\documentclass[aps,prb,reprint,superscriptaddress]{revtex4-1}
\usepackage{graphicx}
\usepackage{amsmath}
\usepackage[dvipsnames,hyperref]{xcolor}
\usepackage{hyperref}
\hypersetup{colorlinks,citecolor=Blue,filecolor=Blue,linkcolor=Blue,urlcolor=Blue}
\usepackage{epstopdf}

\begin{document}
	
\title{Negative \!-\textit{U} and polaronic behavior of the Zn-O divacancy in ZnO}
	
\author{Y. K. Frodason}\email[E-mail: ]{ymirkf@fys.uio.no}
\affiliation{University of Oslo, Centre for Materials Science and Nanotechnology, N-0318 Oslo, Norway}
\author{K. M. Johansen}
\affiliation{University of Oslo, Centre for Materials Science and Nanotechnology, N-0318 Oslo, Norway}
\author{A. Alkauskas}
\affiliation{Center for Physical Sciences and Technology (FTMC), Vilnius LT-10257, Lithuania}	
\author{L. Vines}
\affiliation{University of Oslo, Centre for Materials Science and Nanotechnology, N-0318 Oslo, Norway}

\date{\today}

\begin{abstract}

Hybrid functional calculations reveal the Zn-O divacancy in ZnO, consisting of adjacent Zn and O vacancies, as an electrically active defect exhibiting charge states ranging from $2+$ to $2-$ within the band gap. Notably, the divacancy retains key features of the monovacancies, namely the negative-\textit{U} behavior of the O vacancy, and the polaronic nature of the Zn vacancy. The thermodynamic charge-state transition levels associated with the negative-\textit{U} behavior $\varepsilon$($0$/$2-$), $\varepsilon$($-$/$2-$) and $\varepsilon$($0$/$-$) are predicted to occur at 0.22, 0.42 and 0.02 eV below the conduction band minimum, respectively, resulting in \textit{U} = $-$0.40 eV. These transition levels are moved closer to the conduction band and the magnitude of \textit{U} is lowered compared to the values for the O vacancy. Further, the interaction with hydrogen has been explored, where it is shown that the divacancy can accommodate up to three H atoms. The first two H atoms prefer to terminate O dangling bonds at the Zn vacancy, while the geometrical location of the third depends on the Fermi-level position. The calculated electrical properties of the divacancy are in excellent agreement with those reported for the E4 center observed by deep-level transient spectroscopy, challenging the O vacancy as a candidate for this level.

\end{abstract}
	
\maketitle	
	
\section{\label{sec:introduction}Introduction}
	
The Zn and O vacancy ($V_{\text{Zn}}$ and $V_{\text{O}}$) are among the most widely studied point defects in ZnO, and are frequently invoked to explain experimental results. First-principles defect calculations based on semi-local density functional theory, as well as purposely designed extrapolation techniques to alleviate the band-gap problem of this theory, have been indispensable in elucidating various properties of these defects in the past \cite{Janotti2009,Janotti2007}. More recently, defect calculations based on optimized hybrid functionals have emerged as a viable, albeit computationally demanding, approach to obtain defect energy levels that generally agree well with experimental data \cite{Chen2017,Miceli2018}. For instance, it is now widely accepted that $V_{\text{O}}$ is a deep donor with negative-\textit{U} character \cite{Oba2008,Alkauskas2011,Clark2010,Lany2010a}, i.e., that the energy gain associated with electron pairing at $V_{\text{O}}$ coupled with a large lattice relaxation overcomes the Coulomb repulsion of the two electrons resulting in a net attractive interaction \cite{Anderson1975}. $V_{\text{Zn}}$, on the other hand, is predicted to act as a deep acceptor that can trap up to four holes in polaronic states \cite{Lany2009,Bjorheim2012,Petretto2015,Frodason2017,Lyons2017}. However, experimental verification related to, e.g., the energy level position and optical signature of $V_{\text{O}}$ and $V_{\text{Zn}}$ remains controversial.
		
Several recent experimental studies highlight the close-associate $V_{\text{Zn}}V_{\text{O}}$ pair, hereby referred to as the divacancy, as another important defect in ZnO, especially in processed samples, e.g., after irradiation, annealing or polishing \cite{Makkonen2016,Johansen2016,Holston2016}. In contrast to its isolated constituents, however, first-principles calculations on the divacancy are scarcely available in the literature \cite{Vidya2011,Bang2014,Johansen2016,Chakrabarty2011,Chakrabarty2012,Shin2014}, and predominantly based on semilocal functionals. In the present work, we apply hybrid functional calculations to investigate the properties of the divacancy. Since the divacancy combines a donor with an acceptor, one might intuitively expect a passive and overall neutral pair. Interestingly, our calculations unveil the divacancy as a highly electrically active defect. Moreover, we find that the divacancy retains characteristics of both isolated constituents, namely the negative-\textit{U} property of $V_{\text{O}}$, and the ability of $V_{\text{Zn}}$ to trap holes in polaronic states. Furthermore, we investigate the interaction between divacancies and hydrogen---an omnipresent impurity that is known to occupy $V_{\text{O}}$ and $V_{\text{Zn}}$ in ZnO---and find that the divacancy can accomodate up to three H atoms. Our results are compared with experimental electron paramagnetic resonance (EPR) spectroscopy, deep-level transient spectroscopy (DLTS) and photoluminescence (PL) data.
		
\section{\label{sec:theoretical_framework}Methodology}

Unless specified, all first-principles calculations were based on the generalized Kohn-Sham (KS) theory with the Heyd-Scuseria-Ernzerhof (HSE) \cite{Krukau2006} hybrid functional and the projector augmented wave method \cite{Bloechl1994,Kresse1994,Kresse1999}, as implemented in the {\scriptsize VASP} code \cite{Kresse1993,Kresse1996}. The screening parameter was fixed to the standard value $\omega = 0.2$ \AA$^{-1}$ \cite{Krukau2006}, and the amount of screened Hartree-Fock exchange was set to $\alpha = 37.5 \%$ \cite{Oba2008}. This parametrization of HSE yields a band gap of 3.42 eV, and an accurate description of the structural and electronic properties of wurtzite ZnO \cite{Oba2011}, which is an indispensable requirement for obtaining reliable defect energy levels \cite{Freysoldt2014}.

Defect formation energies and thermodynamic charge-state transition levels were calculated by following a well-established method \cite{Zhang1991,Freysoldt2014}. For instance, the formation energy of the divacancy in charge-state $q$ is given by

\begin{equation}
	E_{\text{f}}^q(V_{\text{Zn}}V_{\text{O}})=E_{\text{tot}}^q(V_{\text{Zn}}V_{\text{O}})-E_{\text{tot}}^{\text{bulk}}+\mu_{\text{Zn}}+\mu_{\text{O}}+q\epsilon_{\text{F}},
\end{equation}

where $E_{\text{tot}}^q(V_{\text{Zn}}V_{\text{O}})$ and $E_{\text{tot}}^{\text{bulk}}$ denote the total energy of the defect-containing and pristine supercells, $\mu_{\text{Zn}}$ and $\mu_{\text{O}}$ is the chemical potential of the removed Zn- and O-atom, and $\epsilon_{\text{F}}$ is the Fermi level position relative to the bulk valence band maximum (VBM). The chemical potential depends on the experimental conditions, but upper and lower bounds are placed by the thermodynamic stability condition $\Delta H^{\text{f}}(\text{ZnO})=\mu_{\text{Zn}}+\mu_{\text{O}}$, where $\Delta H^{\text{f}}(\text{ZnO})$ is the formation enthalpy of ZnO. The upper limit of $\mu_{\text{O}}$ (O-rich conditions) and $\mu_{\text{Zn}}$ (O-poor conditions) is given by the total energy per atom of an O$_{2}$ molecule and metallic Zn, respectively \cite{Janotti2007}. Note that the formation energy of the divacancy is independent of the specific conditions, as $\mu_{\text{Zn}}$ and $\mu_{\text{O}}$ are connected through the stability condition. The chemical potential of H was referenced to H$_{2}$, including H$_{2}$O as a limiting phase under O-rich conditions. For charged defects, we applied the anisotropic \cite{Kumagai2014} Freysoldt-Neugebauer-Van de Walle (FNV) correction to the formation energy \cite{Freysoldt2009,Komsa2012}. 

Thermodynamic charge-state transition levels $\varepsilon(q_{1}/q_{2})$ are given by the Fermi level position for which the formation energy of a defect in two charge-states $q_{1}$ and $q_{2}$ is equal \cite{Freysoldt2014}. The effective correlation energy $U$ for a defect $d$ exhibiting three successive charge-states $q_{1}$, $q_{2}$ and $q_{3}$ is given by the difference between the corresponding thermodynamic charge-state transition levels, i.e., $U=\varepsilon(q_{2}/q_{3})-\varepsilon(q_{1}/q_{2})=E_{\text{f}}^{q_{1}}(d)+E_{\text{f}}^{q_{3}}(d)-2E_{\text{f}}^{q_{2}}(d)$ \cite{Anderson1975,Stoneham1983}. Optical emission and absorption energies, and activation energies for carrier emission, are estimated by using the effective one-dimensional configuration coordinate (CC) model, as described in Refs. \cite{Freysoldt2014,Alkauskas2012,Wickramaratne2018}.
	
Lyons \textit{et al.} \cite{Lyons2017} recently pointed out that a 192-atom supercell is required to ensure converged defect energy levels for $V_{\text{O}}$ in ZnO. After carrying out supercell size-tests for $V_{\text{O}}$ and $V_{\text{Zn}}V_{\text{O}}$, we arrive at the same conclusion. The slow convergence for $V_{\text{O}}$ is mainly caused by: (i) The rather extended defect wave function, which can overlap between neighboring supercells and cause a spurious defect-state dispersion \cite{Freysoldt2014}, and (ii) The large local lattice relaxation associated with its negative-\textit{U} behavior. The former issue can be alleviated by sampling special \textit{k}-points \cite{Walle2004}, but the latter requires an increase in supercell-size. For these reasons, we employ the 192-atom supercell with a plane-wave energy cutoff of 500 eV and a special \textit{k}-point at ($\frac{1}{4}$,$\frac{1}{4}$,$\frac{1}{4}$) for defect calculations. Due to defect-state dispersion, the formation energy of defects involving $V_{\text{O}}$ is found to converge slowly as a function of supercell-size if a $\Gamma$-only \textit{k}-point sampling is used.

\section{\label{sec:results}Results}

\subsection{\label{sec:constituents}Monovacancies}

Before addressing the divacancy, we briefly revisit the electronic properties of the isolated monovacancies. As already mentioned, $V_{\text{O}}$ is a deep double donor exhibiting negative-\textit{U} behavior. The charge-neutral O vacancy induces one fully occupied symmetric KS defect state ($a_1$) inside the band gap, and three empty ones resonant with the conduction band \cite{Janotti2007,Lany2010a}. The negative-\textit{U} behavior is caused by the large difference in local lattice relaxation for the three possible occupations of the $a_1$ state \cite{Janotti2009}. When $a_1$ is fully occupied ($a_1^2$), the vacancy undergoes an inward relaxation in the breathing mode corresponding to $-$10\% of the bulk Zn--O bond length, whereas when $a_1$ is half-filled ($a_1^1$) and empty ($a_1^0$), the relaxation is outward by 6\% and 24\%. The thermodynamic $\varepsilon$($2+$/$0$), $\varepsilon$($+$/$0$) and $\varepsilon$($+$/$2+$) transition levels are predicted to occur at 1.33, 1.57 and 1.08 eV below the conduction band minimum (CBM), respectively, resulting in $U=-0.49$ eV. These results are in good agreement with the aforementioned calculations by Lyons \textit{et al.} \cite{Lyons2017}, and consistent with photo-EPR data on $V_{\text{O}}$ \cite{Evans2008,Wang2009,Lyons2017}.

The doubly negatively charged Zn vacancy introduces four KS defect states in the band gap, all of which are completely filled with electrons \cite{Lany2007}. Upon removal of an electron, the resulting hole is localized at a single nearest-neighbor O ion in the form of a polaron \cite{Galland1970}. Up to four hole polarons can be stabilized at $V_{\text{Zn}}$, resulting in charge states ranging from $2+$ to $2-$ in the band gap \cite{Lany2009,Bjorheim2012,Petretto2015,Frodason2017,Lyons2017}. The calculated $\varepsilon$($2+$/$+$), $\varepsilon$($+$/$0$), $\varepsilon$($0$/$-$) and $\varepsilon$($-$/$2-$) transition levels are positioned at 0.25, 0.89, 1.40 and 1.96 eV above the VBM, respectively \cite{Frodason2017}.

\subsection{\label{sec:electronic-properties}Divacancy}

The divacancy can nominally exist in two different configurations, axial (aligned along the [0001] direction) or azimuthal (lying in the basal plane). Holston \textit{et al.} \cite{Holston2016} have assigned an EPR signal to the azimuthal configuration of the divacancy only, and our calculations indicate that this configuration is more stable, although the difference in energy is small. For this reason, we present results for the azimuthal configuration only.

\subsubsection{Electronic and structural properties, and stability}

\begin{figure}[!htb]
	\includegraphics[width=\columnwidth]{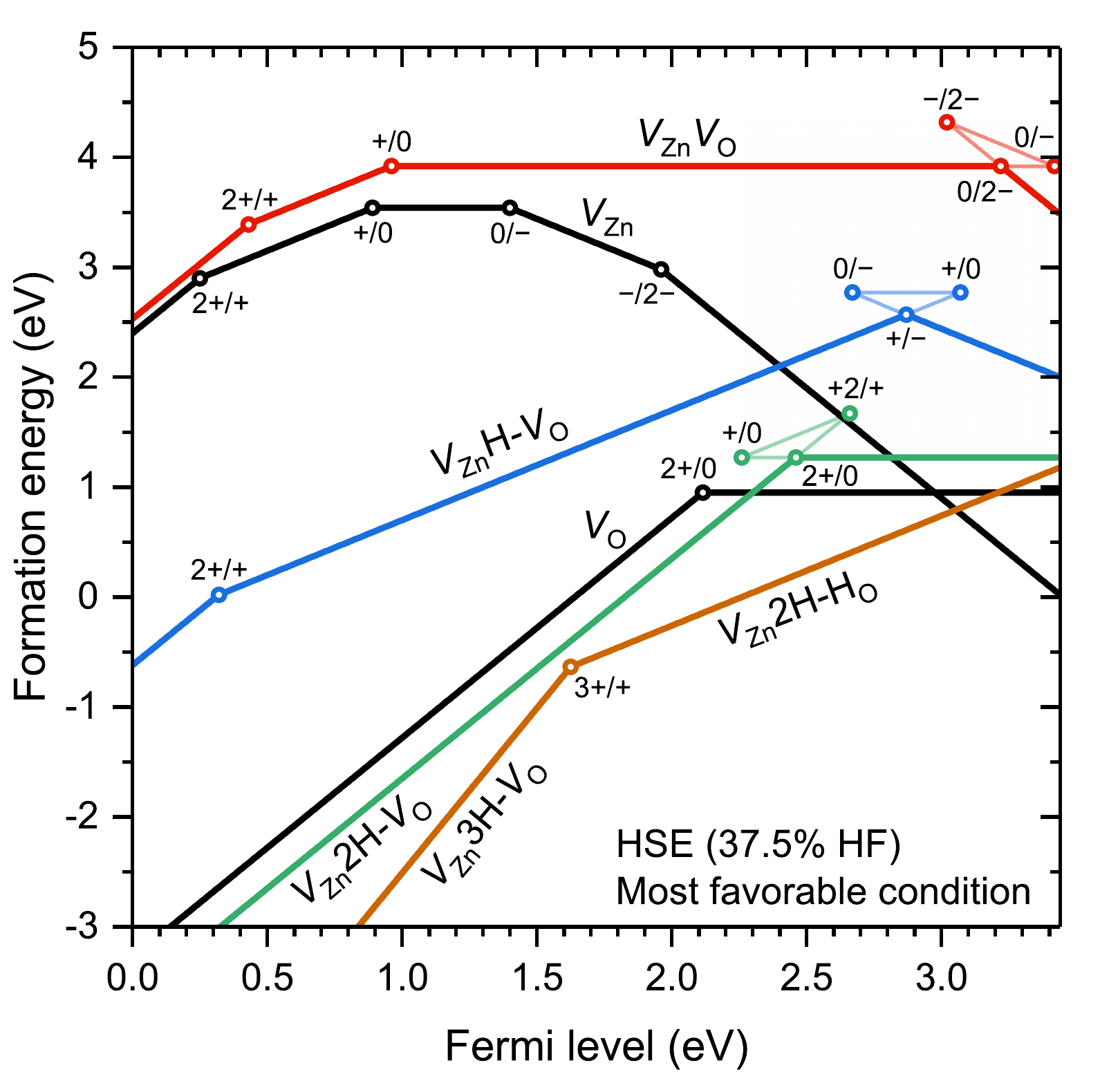}
	\caption{Formation energies as a function of the Fermi level position under the most favorable condition for each defect, i.e., O-rich for $V_{\text{Zn}}$ and O-poor for $V_{\text{O}}$ and $V_{\text{Zn}}V_{\text{O}}n\text{H}$. Formation energies for the opposite limit can be acquired by adding 3.49 eV for $V_{\text{Zn}}$ and $V_{\text{O}}$, and  $n\times$1.48 eV for $V_{\text{Zn}}V_{\text{O}}n\text{H}$.}
	\label{fig:formation-energy}
\end{figure}

\begin{figure}[!htb]
	\includegraphics[width=\columnwidth]{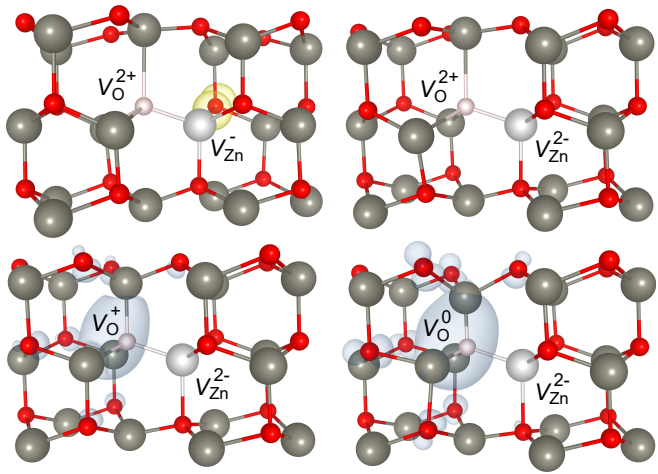}
	\caption{Relaxed divacancy structures. The hole (yellow isosurface) is trapped in a polaronic state at one of the three O ions associated with $V_{\text{Zn}}$. The $a_{1}$ defect state (blue isosurface) of $V_{\text{O}}$ can accommodate two electrons, with a large difference in relaxation for each occupation.}
	\label{fig:divacancy-relaxation}
\end{figure}

Figure \ref{fig:formation-energy} shows the formation energy of $V_{\text{Zn}}$, $V_{\text{O}}$, $V_{\text{Zn}}V_{\text{O}}$ and hydrogenated divacancies with up to three hydrogen atoms as a function of the Fermi level position. Evidently, the divacancy is highly electrically active, displaying charge states ranging from $2+$ to $2-$ in the band gap. Interestingly, it seems to retain certain characteristics of its constituents; the thermodynamic charge-state transition levels in the lower part of the band gap are positioned close to the respective levels of the isolated $V_{\text{Zn}}$, while those in the upper part exhibit negative-\textit{U} characteristics similar to $V_{\text{O}}$. These results can be understood by inspecting the electronic and atomic structure of the divacancy, as shown in Fig. \ref{fig:divacancy-relaxation}. As a double donor, $V_{\text{O}}$ will transfer two electrons to $V_{\text{Zn}}$, thus completely filling its defect states, so the charge-neutral divacancy can be viewed as a $V_{\text{Zn}}^{2-}$ and $V_{\text{O}}^{2+}$ pair. Starting from the neutral charge state, our calculations show that the divacancy can trap a hole polaron at one or two of the three O ions immediately adjacent to $V_{\text{Zn}}$. The resulting $\varepsilon$($2+$/$+$) and $\varepsilon$($+$/$0$) transition levels occur at 0.43 and 0.96 eV above the VBM. The $+3$ charge state (with a hole polaron at each O ion) could not be stabilized.

We also find that the charge-neutral divacancy can capture two electrons in a deep $a_1$ KS defect state (blue isosurfaces in Fig. \ref{fig:divacancy-relaxation}) at $V_{\text{O}}$. Notably, the thermodynamic charge-state transition occurs directly from $0$ to $2-$ \cite{Chakrabarty2011}. This is an important result, because it demonstrates that the isolated $V_{\text{O}}$ is not the only intrinsic defect exhibiting negative-\textit{U} in ZnO, i.e., spectroscopic signatures indicating negative-\textit{U} need not necessarily arise from the isolated $V_{\text{O}}$ only. The position of the $\varepsilon$($0$/$2-$) transition level of the divacancy is shifted strongly up towards the CB, relative to the $\varepsilon$($2+$/$0$) level of $V_{\text{O}}$. This shift is mainly caused by the Coulomb repulsion between $V^{2-}_{\text{Zn}}$ and the electrons occupying the $a_{1}$ state. Again, the negative-\textit{U} behavior results from the large difference in local lattice relaxation around $V_{\text{O}}$ depending on the occupation of $a_{1}$. The breathing-mode relaxation is inward by $-$15\% for $a_{1}^2$, whereas it is outward by 1\% and 21\% for $a_{1}^1$ and $a_{1}^0$, respectively. Note that the relaxation is more inward for the divacancy compared to $V_{\text{O}}$ (Sec. \ref{sec:constituents}). Inward relaxation lowers the energy of the $a_{1}$ state, but increases the strain energy \cite{Janotti2007}. For the divacancy, this energy balance is altered because the number of Zn ions immediately adjacent to $V_{\text{O}}$ is lowered from four to three. The calculated $\varepsilon$($0$/$2-$), $\varepsilon$($-$/$2-$) and $\varepsilon$($0$/$-$) transition levels occur at 0.22, 0.42 and 0.02 eV below the CBM, respectively, resulting in $U=-$0.40 eV, which means the magnitude of \textit{U} is lowered relative to the value for $V_{\text{O}}$.

Holston \textit{et al.} \cite{Holston2016} recently assigned a photo-EPR signal \cite{Leutwein} to the paramagnetic $(V_{\text{Zn}}^{-}V_{\text{O}}^{2+})^{+}$ state ($S=1/2$) in neutron-irradiated ZnO, with the unpaired spin (hole polaron) residing primarily on one of the three O ions immediately adjacent to $V_{\text{Zn}}$. Our prediction is in line with this result, i.e., we find that removing an electron from the neutral divacancy indeed produces a hole polaron at one of the three O ions associated with $V_{\text{Zn}}$ (the hole state corresponds to the yellow isosurface in Fig. \ref{fig:divacancy-relaxation}). The same photo-EPR signal has also been observed in ZnO after 1.7, 2.5 and 3.0 MeV electron irradiation \cite{Schallenberger1976,Soriano}, i.e., for energies above the displacement threshold of both O (310 keV) and Zn (900 keV) in ZnO \cite{Locker1972}. 

Although $(V_{\text{Zn}}V_{\text{O}})^{-}$ is thermodynamically unstable, it might be possible to create it in a metastable manner, e.g., by optical excitation. In fact, the analogous paramagnetic isolated $V_{\text{O}}^{+}$ state is detectable at low temperatures by photo-EPR \cite{Evans2008,Wang2009}. Holston \textit{et al.} \cite{Holston2016} initially envisioned the paramagnetic $(V_{\text{Zn}}^{2-}V_{\text{O}}^{+})^{-}$ state as responsible for the aforementioned divacancy photo-EPR signal. However, this model was discarded as its EPR spectrum would have negative g shifts and well resolved hyperfine interactions with the adjacent $^{67}$Zn nuclei, similar to the isolated $V_{\text{O}}^{+}$ \cite{Holston2016}. Moreover, seeing as the Fermi level was lowered by the neutron-irradiation \cite{Holston2016}, the divacancies were most likely charge-neutral before illumination.

A pertinent question is whether divacancies are likely to form, and if they are stable at room temperature. Since $V_{\text{Zn}}$ and $V_{\text{O}}$ favor opposite conditions, the formation energy of $V_{\text{Zn}}V_{\text{O}}$ is high, indicating a low equilibrium concentration. Indeed, experimentally, divacancies are typically observed in ZnO after post-growth processing, e.g., after irradiation, ion implantation, annealing or polishing \cite{Makkonen2016,Johansen2016,Holston2016}. Regarding stability, we find that the defect reaction $V_{\text{Zn}}^{2-} + V_{\text{O}}^{0} \rightarrow (V_{\text{Zn}}V_{\text{O}})^{2-}$ lowers the total energy by 0.83 eV under \textit{n}-type conditions, which means that the divacancy will be stable at room temperature. Using the climbing nudged elastic band method \cite{Henkelman2000} and the Perdew-Burke-Ernzerhof (PBE) \cite{Perdew1996} functional, we have also investigated migration of the divacancy, i.e., nearest-neighbor Zn and O atoms hopping into the vacant Zn and O sites, respectively. The resulting migration barrier is about 1.2 eV, which is lower than those predicted previously for monovacancies in ZnO \cite{Janotti2007}. Thus, divacancies may diffuse at relatively low temperatures and form larger vacancy clusters \cite{Bang2014} or other complexes \cite{Holston2016}. Indeed, formation of vacancy clusters in ZnO has been reported in several experimental studies \cite{Makkonen2016,Johansen2016,Chan2013}. 

\subsubsection{\label{sec:electron-emission}Dynamics of electron emission and capture}

\begin{figure}[!htb]
	\includegraphics[width=\columnwidth]{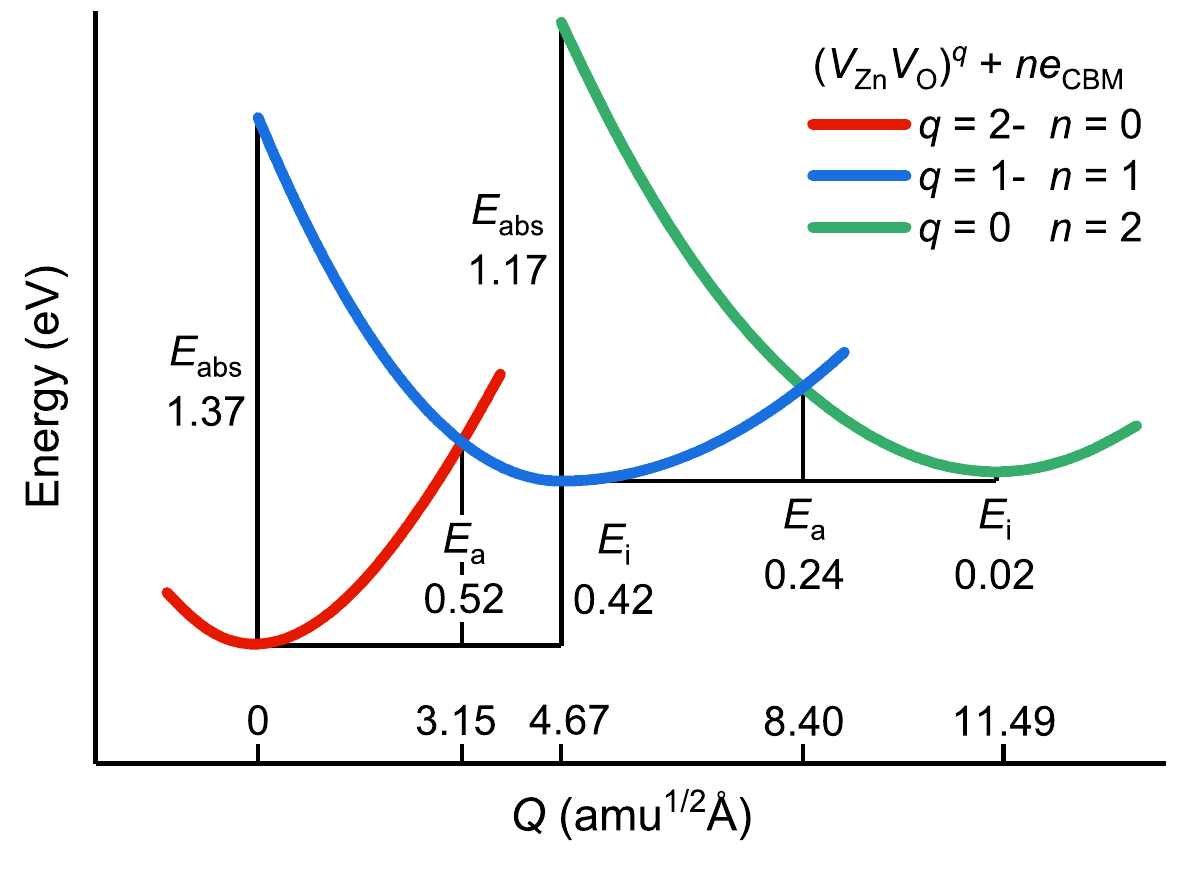}
	\caption{CC diagram describing the dynamics of electron emission and capture between the divacancy and the CBM. Optical absorption ($E_{\text{abs}}$), ionization ($E_{\text{i}}$) and activation ($E_{\text{a}}$) energies for electron emission are provided (all in eV). The magnitude of Q is given at the minima of the potential energy curves and the crossing points between them.}
	\label{fig:divacancy-cc}
\end{figure}

In order to allow comparison of our results with experimental data, e.g., energy level positions obtained by DLTS, we have constructed a CC diagram (shown in Fig. \ref{fig:divacancy-cc}) describing the two-step process of electron emission from the doubly negatively charged divacancy to the CBM. The generalized coordinate (Q) along the horizontal axis corresponds to the mass-weighted atomic displacement between the equilibrium structure of $V_{\text{Zn}}V_{\text{O}}$ for the three different charge states, namely $2-$, $-$ and $0$ \cite{Alkauskas2012}, and can be loosely viewed as the magnitude of outward-breathing relaxation around $V_{\text{O}}$. The total energy as a function of Q results in a potential energy curve for each charge state, and the energy is minimized at the equilibrium structure for each curve. The potential energy curves are displaced in energy along the vertical axis by the Fermi level position of the thermodynamic charge-state transition levels relative to the CBM, i.e., the difference in energy between the states in their equilibrium configurations. We will refer to these differences in energy as defect ionization energies ($E_{\text{i}}$) \cite{Wickramaratne2018}.

As shown in Fig. \ref{fig:divacancy-cc}, we obtain an activation energy ($E_{\text{a}}$) of 0.52 eV for electron emission from $(V_{\text{Zn}}V_{\text{O}})^{2-}$ to the CBM. This activation energy corresponds to the sum of $E_{\text{i}}$ and the capture barrier of 0.10 eV, which is taken as the energy required to reach the crossing point between the curves for $(V_{\text{Zn}}V_{\text{O}})^{-}+e^{-}_{\text{CBM}}$ and $(V_{\text{Zn}}V_{\text{O}})^{2-}$. Keep in mind that this barrier should be viewed as an upper estimate \cite{Mooney1999,Wickramaratne2018}. Moreover, the temperature effects described in Ref. \cite{Wickramaratne2018} are also neglected here. For emission from $(V_{\text{Zn}}V_{\text{O}})^{-}$, we obtain a lower activation energy of 0.24 eV for emission and a capture barrier of 0.22 eV. 

\subsubsection{\label{sec:optical-properties}Optical transitions}

Several broad luminescence bands have been observed in the visible part of the emission spectrum of ZnO, and there is a plethora of studies linking these to various point defects, including $V_{\text{Zn}}$ and $V_{\text{O}}$. However, consensus is lacking and the defect origin of most bands remains unknown. Previous hybrid functional calculations suggest that the isolated vacancies are unlikely to give rise to luminescence in the visible range under \textit{n}-type conditions \cite{Frodason2017,Lyons2017}. However, the optical properties of the divacancy could be very different from those of $V_{\text{Zn}}$ and $V_{\text{O}}$ \cite{Lyons2017,Frodason2018}. 

\begin{figure}[!htb]
	\includegraphics[width=\columnwidth]{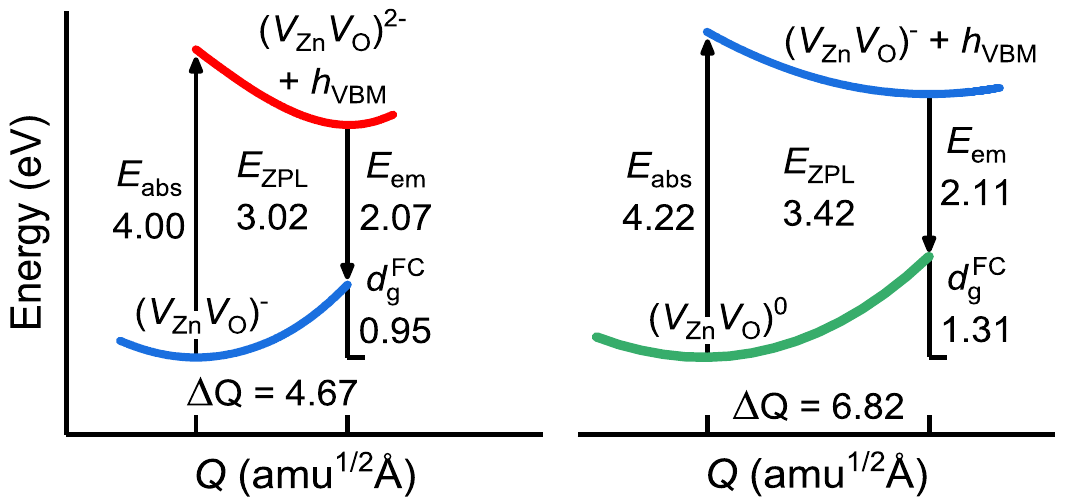}
	\caption{CC diagram for optical transitions involving capture of a hole at the VBM by $(V_{\text{Zn}}V_{\text{O}})^{2-}$ and $(V_{\text{Zn}}V_{\text{O}})^{-}$. Zero phonon line ($E_{\text{ZPL}}$), emission ($E_{\text{em}}$), absorption ($E_{\text{abs}}$) and ground state relaxation ($d_{\text{g}}^{\text{FC}}$) energies (all in eV), and total mass-weighted distortion (in amu$^{1/2}$\AA) is provided.}
	\label{fig:divacancy-cc-pl}
\end{figure}

We have considered three different optical transitions for the divacancy, namely hole capture from the VBM by $(V_{\text{Zn}}V_{\text{O}})^{2-}$ and $(V_{\text{Zn}}V_{\text{O}})^{-}$, and electron capture from the CBM by $(V_{\text{Zn}}V_{\text{O}})^{+}$. Figure \ref{fig:divacancy-cc-pl} shows the calculated CC diagrams for the two former transitions. Again, the potential energy curves in the CC diagrams are vertically displaced in energy by the Fermi level position of the respective thermodynamic charge-state transition levels relative to the VBM. In the context of luminescence, this is usually referred to as the zero phonon line energy, $E_{\text{ZPL}}$. In the Franck-Condon approximation, optical transitions take place without atomic motion, i.e., emission ($E_{\text{em}}$) and absorption ($E_{\text{abs}}$) energies are given by the vertical arrows in Fig. \ref{fig:divacancy-cc-pl}. After an optical transition, the defect will relax to its equilibrium configuration by emitting phonons, losing the Franck-Condon relaxation energy (denoted $d_{\text{g}}^{\text{FC}}$ for the ground state).

Interestingly, the difference in $E_{\text{ZPL}}$ for the transitions in Fig. \ref{fig:divacancy-cc-pl} is offset by a difference in $d_{\text{g}}^{\text{FC}}$, resulting in close emission energies of 2.07 and 2.11 eV. The capture of an electron at the CBM by the positively charged divacancy results in a lower emission energy of 1.66 eV; the CC diagram is not shown here, but is similar to what has been found previously for other $V_{\text{Zn}}$ related defects \cite{Frodason2017,Lyons2017,Frodason2018}. Overall, our results show that the divacancy can give rise to broad luminescence peaking in the 1.6--2.1 eV range. We also note that both considered transitions should be dipole allowed. In the case of radiative electron capture the CB electrons have mostly Zn 4\textit{s} character, while the final states have mostly O 2\textit{p} \cite{Freysoldt2014} character. For radiative hole capture the initial state is a perturbed VB state which is composed mainly of O 2\textit{p} states, while the defect state has mostly Zn 4\textit{s} character. A different parity ensures a strong transition dipole moment and thus dipole-allowed transitions.
 Experimental studies on the optical properties of divacancies are scarce in the literature, but our calculations are consistent with the results reported by Dong et al. \cite{Dong2010}. Based on cathodoluminescence and PAS data, Dong et al. \cite{Dong2010} suggested that the emission from large vacancy clusters peak at up to 2.1 eV, which shifts to lower energies with decreasing cluster size. 

\subsection{\label{sec:hydrogenated-divacancy}Hydrogen decorated divacancy}

Besides being a ubiquitous impurity, hydrogen strongly influences the electrical and optical properties of ZnO, notably as a source of unintentional \textit{n}-type conductivity \cite{Mollwo1954,Thomas1956,Walle2000}. Interstitial hydrogen (H$_{\text{i}}$) acts exclusively as a donor by forming a strong O--H bond, preferably at the axial bond-centered site \cite{Walle2000,Wardle2005,Lavrov2009}. However, H$_{\text{i}}$ is mobile at room temperature \cite{Wardle2006,Johansen2008,Hupfer2017}, and is thus likely to become trapped at defects. Importantly, the Coulomb attraction between H$^{+}_{\text{i}}$ and $V_{\text{Zn}}^{2-}$ results in highly stable $V_{\text{Zn}}n\text{H}$ complexes with $n=1,2,3$, wherein H remains a donor by terminating O dangling bonds; the first two H atoms successively passivate the acceptor levels of $V_{\text{Zn}}$, whereas the third transforms the complex into a shallow single donor. H can also be trapped at $V_{\text{O}}$, but in this case the hydrogen behaves as an acceptor (H$^-$) substituting for oxygen (fourfold coordinated). The resulting complex acts as a single shallow donor (H$^{+}_{\text{O}}$) \cite{Walle2007,Lavrov2009}.

Seeing as both $V_{\text{Zn}}$ and $V_{\text{O}}$ present potential trapping sites for H in the divacancy, different configurations were explored. Furthermore, to assess the divacancy as a hydrogen trap, removal energies were calculated as the difference in formation energy between the hydrogenated divacancy and the two remaining defects (calculated within separate supercells) when one H is removed and placed in its most stable isolated configuration ($\text{H}_{\text{i}}^{+}$).

\begin{figure}[!htb]
	\includegraphics[width=\columnwidth]{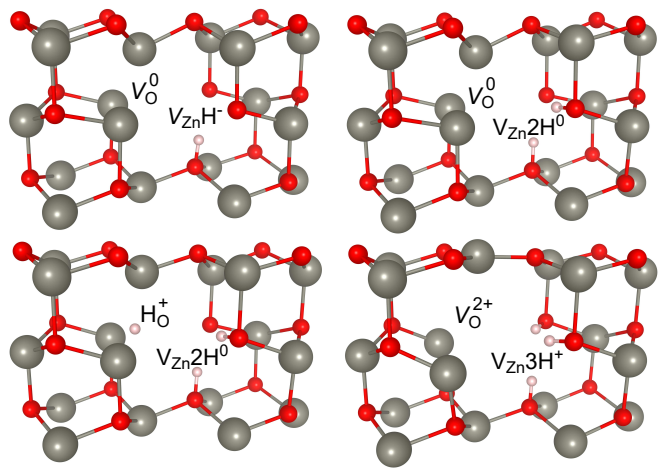}
	\caption{Relaxed structures for the divacancy containing up to three H atoms. The first two H atoms terminate O dangling bonds, either at the bond centered axial or azimuthal site. Depending on the Fermi level position, the third H atom will occupy $V_{\text{O}}$ or terminate the final O dangling bond.}
	\label{fig:divacancy-relaxation-H}
\end{figure}

As shown in Fig. \ref{fig:divacancy-relaxation-H}, the first two H atoms prefer to terminate O dangling bonds at $V_{\text{Zn}}$, resulting in $V_{\text{Zn}}\text{H--}V_{\text{O}}$ and $V_{\text{Zn}}2\text{H--}V_{\text{O}}$ complexes. The formation of these complexes lowers the total energy substantially, as evidenced by the large respective H removal energies of 2.84 and 2.02 eV. These are only slightly lower than for the $V_{\text{Zn}}\text{H}$ and $V_{\text{Zn}}2\text{H}$ complexes \cite{Lyons2017}. For comparison, removal energies of 2.08 and 1.26 eV are obtained for the $V_{\text{Zn}}\text{--H}_{\text{O}}$ and $V_{\text{Zn}}\text{H--H}_{\text{O}}$ configurations. As indicated in Fig. \ref{fig:formation-energy}, the location of the third H atom depends on the position of the Fermi level. When $\epsilon_{\text{F}}>1.63$ eV, the H atom will preferentially occupy $V_{\text{O}}$, resulting in the $(V_{\text{Zn}}2\text{H--H}_{\text{O}})^{+}$ complex with a calculated H removal energy of 1.26 eV. When $\epsilon_{\text{F}}<1.63$ eV, the H atom will instead terminate the final dangling bond at $V_{\text{Zn}}$, resulting in $(V_{\text{Zn}}3\text{H--}V_{\text{O}})^{3+}$. However, this latter configuration is unstable, as the H removal energy is $-$0.48 eV.

As can be seen in Fig. \ref{fig:formation-energy}, the negative-\textit{U} behavior persists for both $V_{\text{Zn}}\text{H--}V_{\text{O}}$ and $V_{\text{Zn}}2\text{H--}V_{\text{O}}$. However, the corresponding transition levels are shifted down in Fermi level position. Indeed, as the acceptor levels of $V_{\text{Zn}}$ are passivated by H, the Coulomb repulsion experienced by the electrons occupying the $a_{1}$ defect state is successively lowered, and the transition levels shift down towards the VB by $\sim$0.4 eV per H atom. Specifically, the $\varepsilon$($-$/$+$) and $\varepsilon$($2+$/$0$) transition levels of $V_{\text{Zn}}\text{H--}V_{\text{O}}$ and $V_{\text{Zn}}2\text{H--}V_{\text{O}}$ are predicted to occur at 0.57 and 0.98 eV below the CBM. The magnitude of \textit{U} is unchanged with respect to the isolated divacancy ($U=-$0.40 eV), which again indicates that it is mainly determined by the number of Zn atoms immediately adjacent to $V_{\text{O}}$. Note that these trends hold for the $\varepsilon$($3+$/$+$) transition of the unstable $V_{\text{Zn}}3\text{H--}V_{\text{O}}$ complex as well. Lastly, the $V_{\text{Zn}}2\text{H--H}_{\text{O}}$ complex behaves like an effective-mass donor, and should be viewed as a $(V_{\text{Zn}}2\text{H})^{0}$ and H$^{+}_{\text{O}}$ pair. At this point, the divacancy is fully saturated with H atoms.

\subsection{\label{sec:E4}Comparison with DLTS and the E4 center}

As previously discussed, the divacancy is not expected to be present in substantial amount in as grown samples, but may be formed during processing. Several electrically active defect centers having energy levels in the upper part of the band gap have been reported in \textit{n}-type ZnO \cite{Monakhov2009}. In this section, we compare our results for the divacancy with experimental DLTS data. We start by envisioning the electron emission from $(V_{\text{Zn}}V_{\text{O}})^{2-}$ during a conventional DLTS measurement for \textit{n}-type material. After a zero-bias filling pulse at low temperature, all divacancies will be filled with electrons. Subsequently, at some elevated temperature and under reverse bias, the first electron will be thermally emitted from $(V_{\text{Zn}}V_{\text{O}})^{2-}$ with $E_{\text{a}}$ = 0.52 eV. Since the electron at $(V_{\text{Zn}}V_{\text{O}})^{-}$ is bound less strongly ($E_{\text{a}}$ = 0.24 eV), emission of the second electron will proceed immediately after the first electron, leaving the defect in the $(V_{\text{Zn}}V_{\text{O}})^{0}$ state. For this reason, the divacancy will emit always two electrons during a conventional DLTS measurement, resulting in a single peak with $E_{\text{a}}$ = 0.52 eV. Indeed, such two-electron emission is common for centers where the electron binding energy shows an inverted order \cite{Hemmingsson1998,Son2012}. 

It is sometimes possible to obtain the activation energy associated with one-electron emission from a negative-\textit{U} center by using DLTS with a short filling pulse and illumination. If the filling pulse is sufficiently short, most charge-neutral divacancies will be prevented from capturing more than one electron.  A new DLTS peak, corresponding to the one-electron emission with $E_{\text{a}}$ = 0.24 eV, may then be observed at a lower temperature than for the two-electron emission. However, even with a short filling pulse, some divacancies may still capture a second electron, which means that they will be frozen out in the temperature range where the one-electron emission peak can be observed, and the peak will decrease to zero over time due to the repetitive pulses required by DLTS \cite{Hemmingsson1998}. This can be avoided by optically ``emptying'' any divacancies having captured two electrons in the preceeding pulse before each filling pulse. This technique has been used, e.g., to observe one-electron emission from the negative-\textit{U} center Z$_{1/2}$ in 4H-SiC \cite{Hemmingsson1998}.

Interestingly, the calculated $E_{\text{a}}$ = 0.52 eV for $V_{\text{Zn}}V_{\text{O}}$ is close to the experimental activation energy of $\sim$0.55 eV for the E4 center in ZnO, which has been observed by several groups and is commonly assigned to a $V_{\text{O}}$-related defect exhibiting negative-\textit{U} \cite{Auret2001,Monakhov2009}. However, theoretical predictions and experimental photo-EPR data \cite{Evans2008,Wang2009,Lyons2017} suggest that the isolated $V_{\text{O}}$ is substantially deeper than the E4 center, and this discrepancy has been a source for a long-standing debate in the community. 

Our theoretical predictions for the divacancy conform to many of the experimental findings on the E4 center: (i) Negative-\textit{U} behavior has been reported for E4 \cite{Frank2007,Ellguth2011}. In particular, by using a short filling pulse and illumination as described above, an apparent energy level of $\sim$0.14 eV has been obtained for the one-electron emission from E4 \cite{Hofmann2007,Frank2007}, which is close to our calculated value of 0.24 eV. (ii) Based on the temperature dependence of the rate of electron capture by E4 during the zero-bias filling pulse, Hupfer \textit{et al.} \cite{Hupfer2016} obtained an experimental capture barrier of $\sim$0.15 eV, in good agreement with our calculated value of 0.22 eV for the divacancy. (iii) E4 exhibits a low concentration in as-grown material, but is prominent after He$^{+}$ and H$^{+}$ irradiation \cite{Auret2001,Hayes2007,Hayes2007a,Hupfer2016}. Its concentration hinges linearly on the ion dose, but amounts to only $\sim$0.23\% of the total vacancy generation obtained from Monte Carlo simulations using the {\scriptsize SRIM} code \cite{Monakhov2009,Ziegler2010}. Such a low generation rate is uncommon for a primary intrinsic defect, even if strong dynamic annealing is taken into account, and favors E4 as a high-order intrinsic defect such as the divacancy \cite{Monakhov2009}. For comparison, the generation rate is about 1-2\% for the divacancy in Si after high energy ion irradiation \cite{Monakhov2009}. E4 has also been observed after 2 MeV electron irradiation \cite{Frank2007}, i.e., above the displacement threshold of both O and Zn in ZnO. (iv) After H$^{+}$ implantation at room temperature, the E4 center anneals out at a rate that fits well with a model invoking migration of H$_{\text{i}}$ and subsequent reaction with E4 \cite{Hupfer2016}; our results show that the divacancy has a strong affinity for H, similar to that of $V_{\text{Zn}}$. Furthermore, E4 is accompanied by a second trap labelled Ep2 after H$^{+}$ irradiation, which similarly exhibits a low introduction rate \cite{Auret2001}. Indeed, Auret \textit{et al.} \cite{Auret2001} suggested that Ep1 (E4) and Ep2 might be high-order rather than primary intrinsic defects. Finally, Ep2 has an apparent energy level at $\sim$0.78 eV, which incidentally fits well with our calculated ionization energy of 0.77 eV for $V_{\text{Zn}}\text{H--}V_{\text{O}}$. 

\section{Conclusion}

Using hybrid functional calculations, we have investigated the divacancy in ZnO. The divacancy is predicted to be a highly electrically active defect, exhibiting charge states ranging from $2+$ to $2-$ in the band gap. Interestingly, it retains both the negative-\textit{U} behavior of $V_{\text{O}}$ and the polaronic nature of $V_{\text{Zn}}$. The former result is of particular interest, as it demonstrates that the isolated $V_{\text{O}}$ is not the only intrinsic defect exhibiting negative-\textit{U} in ZnO. We have also studied the interaction between divacancies and H, and find that the divacancy can accommodate up to three H atoms, the first two of which prefer to terminate O dangling bonds at $V_{\text{Zn}}$. Based on comparison of our results with experimental DLTS data in the literature \cite{Monakhov2009,Auret2001,Hayes2007,Hayes2007a,Ellguth2011,Frank2007}, $V_{\text{Zn}}V_{\text{O}}$ is proposed as a potential origin of the E4 center, which is commonly associated with a $V_{\text{O}}$-related defect with negative-\textit{U}.

\begin{acknowledgments}
	
Financial support was kindly provided by the Research Council of Norway and University of Oslo through the frontier research project FUNDAMeNT (no. 251131, FriPro ToppForsk-program). A.A. was supported by Marie Sk\l{}odowska-Curie Action of the European Union (project \textsc{Nitride}-SRH, Grant No. 657054). The computations were performed on resources provided by UNINETT Sigma2 - the National Infrastructure for High Performance Computing and Data Storage in Norway.
	
\end{acknowledgments}
	
\bibliographystyle{apsrev4-2}
\bibliography{main-text-with-figures}	

\end{document}